\documentclass[3p,times]{elsarticle}

\usepackage{ecrc}

\volume{00}

\firstpage{1}

\journalname{Nuclear Physics A}

\runauth{Takayasu Sekihara et al.}

\jid{NUPHA}

\jnltitlelogo{Nuclear Physics A}

\CopyrightLine{2012}{Published by Elsevier Ltd.}

\usepackage{amssymb}

\usepackage[figuresright]{rotating}
\usepackage{bm}
\usepackage{amsmath, amssymb}

\begin{document}

\begin{frontmatter}

\dochead{}



\title{Mesonic and non-mesonic branching ratios of $K^{-}$ absorption
  in the nuclear medium}

\author[a]{Takayasu Sekihara}
\author[a]{Junko Yamagata-Sekihara}
\author[b]{Daisuke Jido}
\author[c]{Yoshiko Kanada-En'yo}

\address[a]{Institute of Particle and
  Nuclear Studies, High Energy Accelerator Research Organization
  (KEK), 1-1, Oho, Ibaraki 305-0801, Japan}

\address[b]{Yukawa Institute for Theoretical Physics, 
  Kyoto University, Kyoto 606-8502, Japan}

\address[c]{Department of Physics, Kyoto University, Kyoto 606-8502, Japan}

\begin{abstract}
  The branching ratios of $K^{-}$ absorption at rest in nuclear matter
  are evaluated from the $K^{-}$ self-energy by using the chiral
  unitary approach for the $s$-wave $\bar{K}N$ amplitude.  We find
  that both the mesonic and non-mesonic absorption potentials are
  dominated by the $\Lambda (1405)$ contributions.  We also observe
  that the mesonic absorption ratio $[\pi ^{-} \Sigma ^{+}] / [\pi
  ^{+} \Sigma ^{-}]$ increases as a function of nuclear density due to
  the interference between $\Lambda (1405)$ and the $I=1$ non-resonant
  background, which is consistent with experimental results.  
  The fraction of the non-mesonic absorption is evaluated to be about
  $30 \%$ at the saturation density.  The branching ratios of the
  $K^{-}$ absorption at rest into deuteron and ${}^{4} \text{He}$ are
  also calculated.
\end{abstract}


\end{frontmatter}


\section{Introduction}
\label{sec1}

Antikaon ($\bar{K}$)-nucleus bound systems have attracted continuous
attention because they are important tools to study the low-energy
$\bar{K} N$ and $\bar{K}$-nucleus interactions and the in-medium
properties of $\bar{K}$~\cite{Batty:1997zp, Friedman:2007zza}.  Kaonic
atoms, which are Coulombic bound states of $K^{-}$-nucleus with the
influence of strong interaction, provide us with unique information on
strong interaction between $K^{-}$ and nucleus at low-energy from
their binding energies and decay widths~\cite{Batty:1997zp,
  Friedman:2007zza}.  In addition, strongly attractive $\bar{K} N$
interaction in the $I=0$ channel, which dynamically generates $\Lambda
(1405)$ as a $\bar{K}N$ quasi-bound state~\cite{Dalitz:1967fp} (see
also~\cite{Hyodo:2011ur}), stimulates recent studies on kaonic nuclei,
which are the $\bar{K}$ few-nucleon systems bound mainly by the strong
interaction~\cite{Friedman:2007zza, Kishimoto:1999yj}. Searches for
kaonic nuclei have been done in recent
experiments~\cite{Agnello:2005qj}, although there is no clear
evidence.

One of the important properties of the $\bar{K}$-nucleus systems to
extract information on the $\bar{K} N$ and $\bar{K}$-nucleus
interactions is their decay patterns.  Branching ratios of stopped
$K^{-}$ absorption were investigated experimentally as early as in the
1970s~\cite{Tovee:1971ga, Veirs:1970fs, Katz:1970ng} and extended in
recent studies, {\it e.g.}, Ref.~\cite{Agnello:2011iq}.  As a result,
we have observed the increase of the absorption ratio $R_{+-}\equiv
[\pi ^{-} \Sigma ^{+}] / [\pi ^{+} \Sigma ^{-}]$ for stopped $K^{-}$
on light nuclei as a function of nuclear number $A$, which implies
sub-threshold modification of the $\bar{K} N$ interaction, in
agreement with earlier calculations~\cite{Staronski:1987he}, and have
determined non-mesonic fractions of stopped $K^{-}$ absorption in
various nuclei.

In order to interpret these branching ratios, we investigate
theoretically the mesonic and non-mesonic branching ratios for stopped
$K^{-}$ absorption in nuclear matter by using a $\bar{K} N$
interaction model as an input~\cite{Sekihara:2012wj}.  In this study
the branching ratios are evaluated from the imaginary part of the
$K^{-}$ self-energy as functions of nuclear density as well as kaonic
energy and momentum.  Actually there are several preceding studies on
the in-medium $K^{-}$ self-energy~\cite{Waas:1997pe}.  Here we
concentrate on the imaginary part of the self-energy to evaluate the
branching ratios of the $K^{-}$-nuclear systems.

\section{Mesonic and non-mesonic branching ratios of $K^{-}$ absorption}
\label{sec2}

\begin{figure}[t]
  \centering
  \begin{tabular*}{\textwidth}{@{\extracolsep{\fill}}ccccc}
    \includegraphics[scale=0.11]{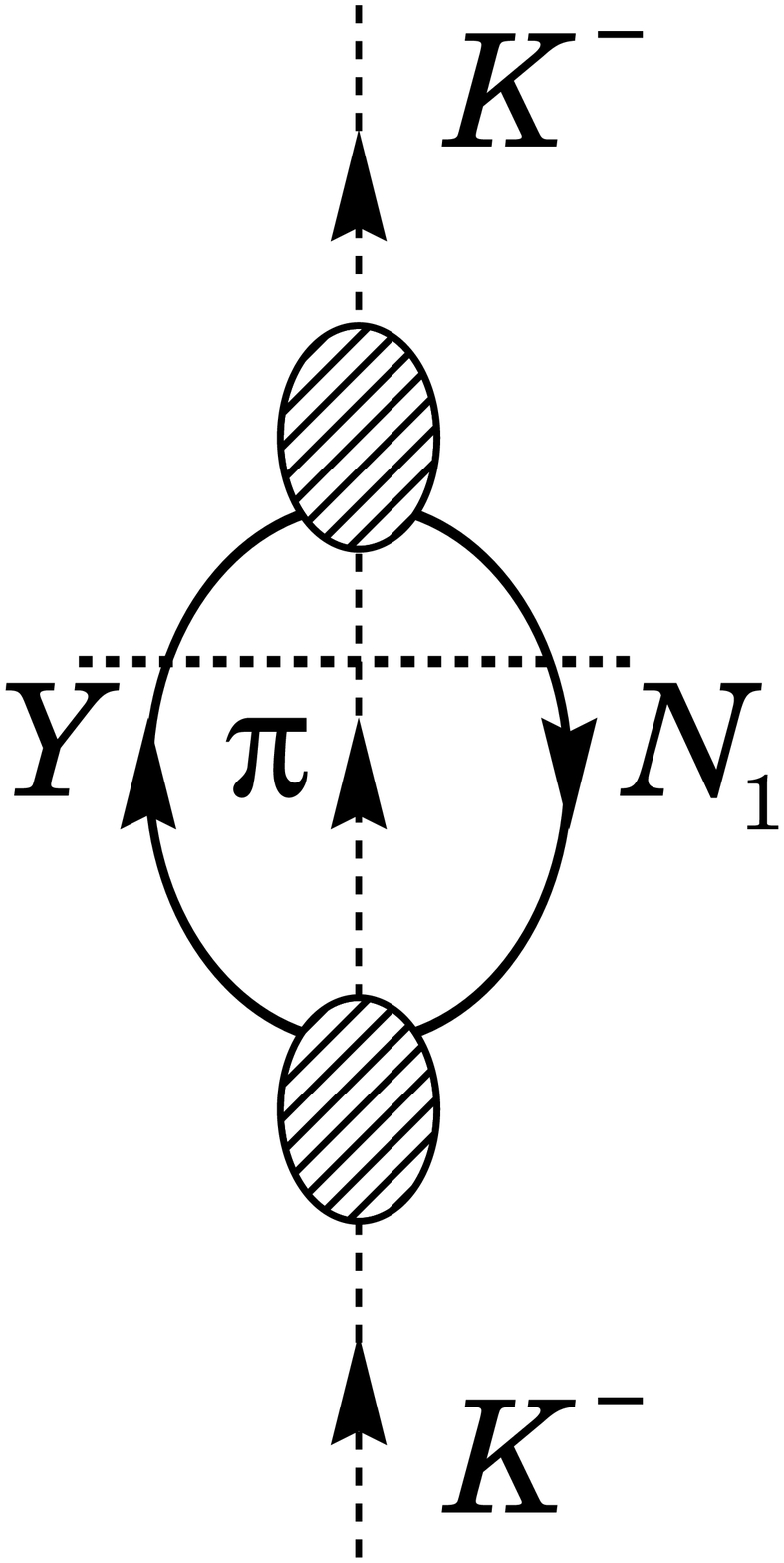}
    &
    \includegraphics[scale=0.11]{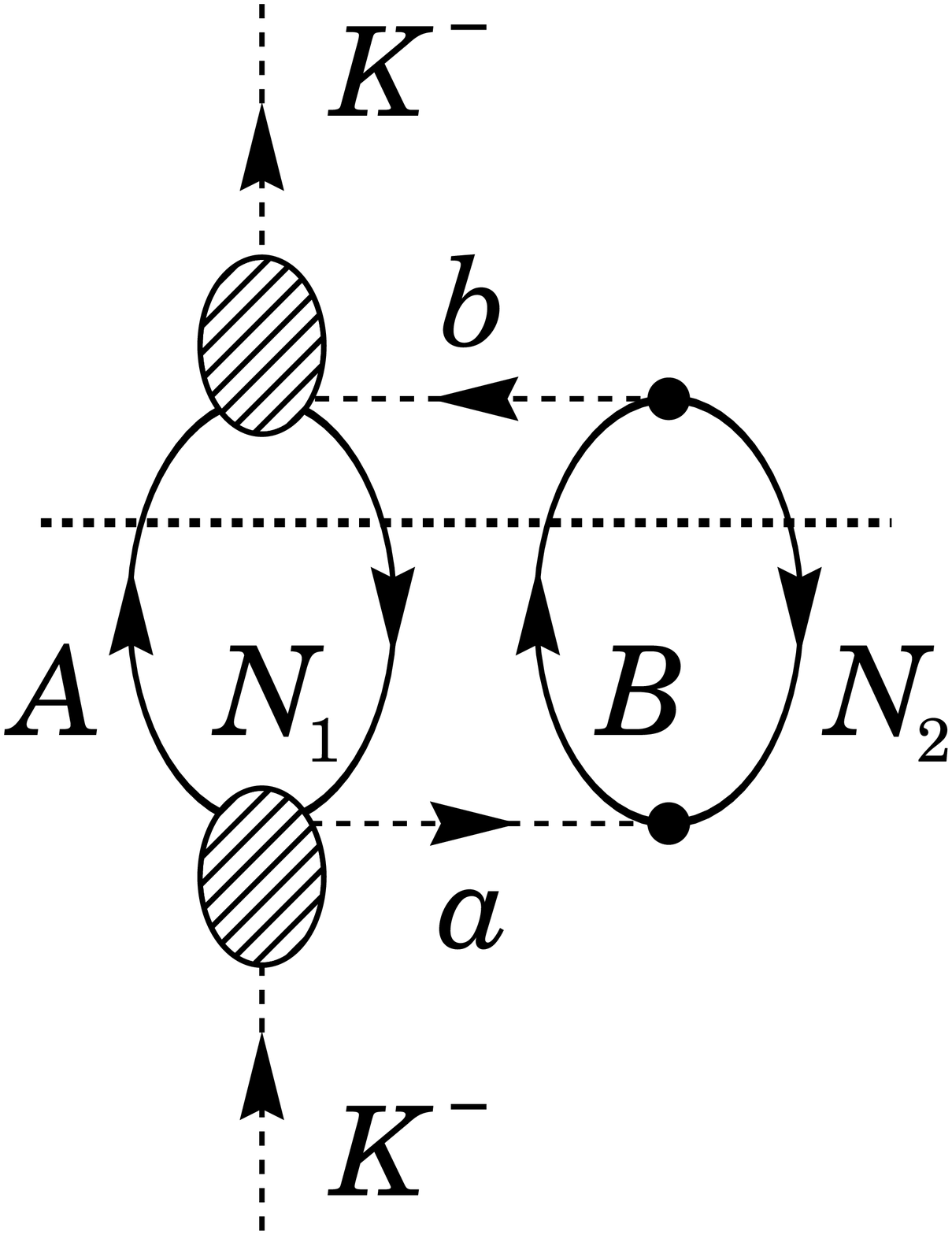}
    &
    \includegraphics[scale=0.11]{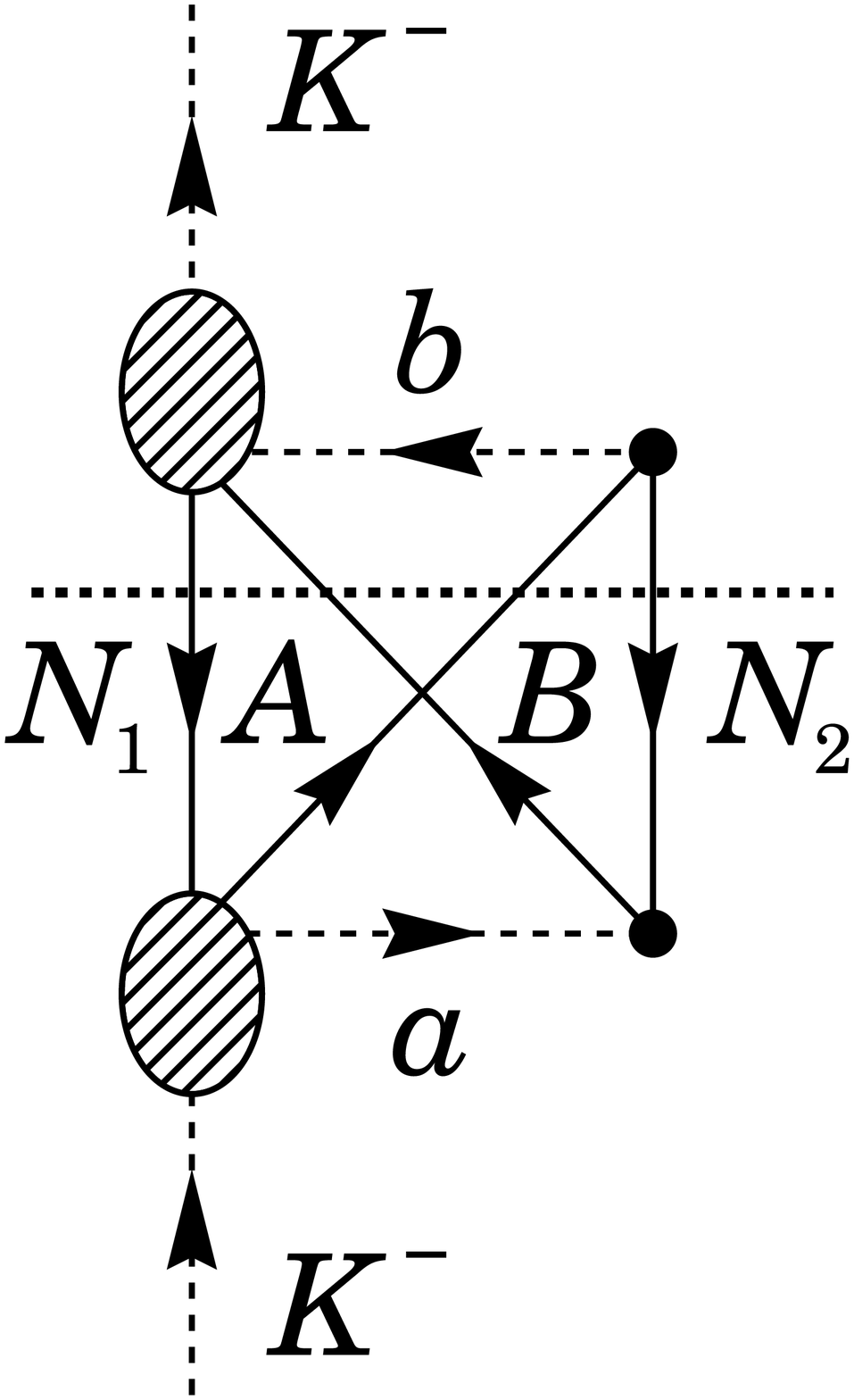}
    &
    \includegraphics[scale=0.11]{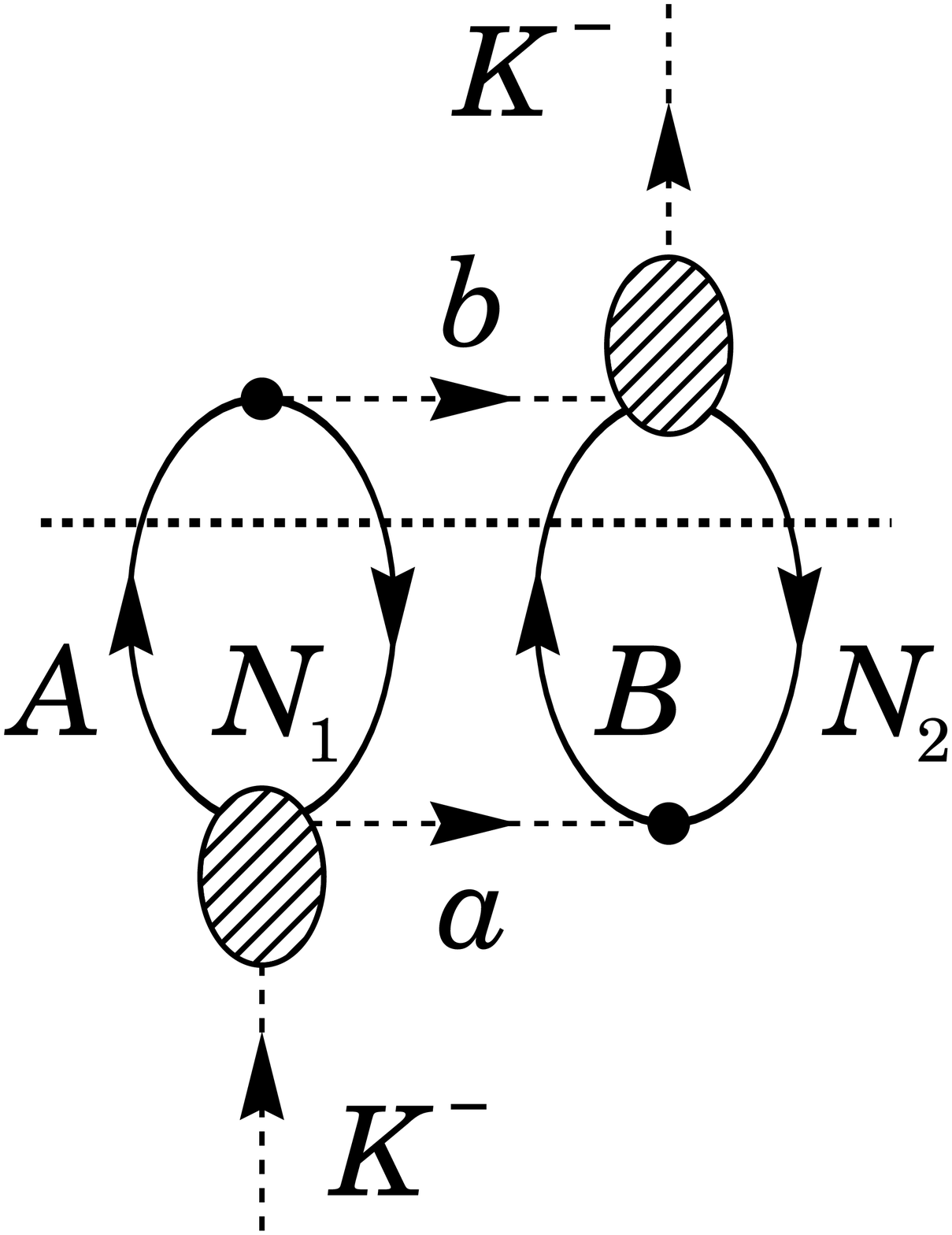}
    & 
    \includegraphics[scale=0.11]{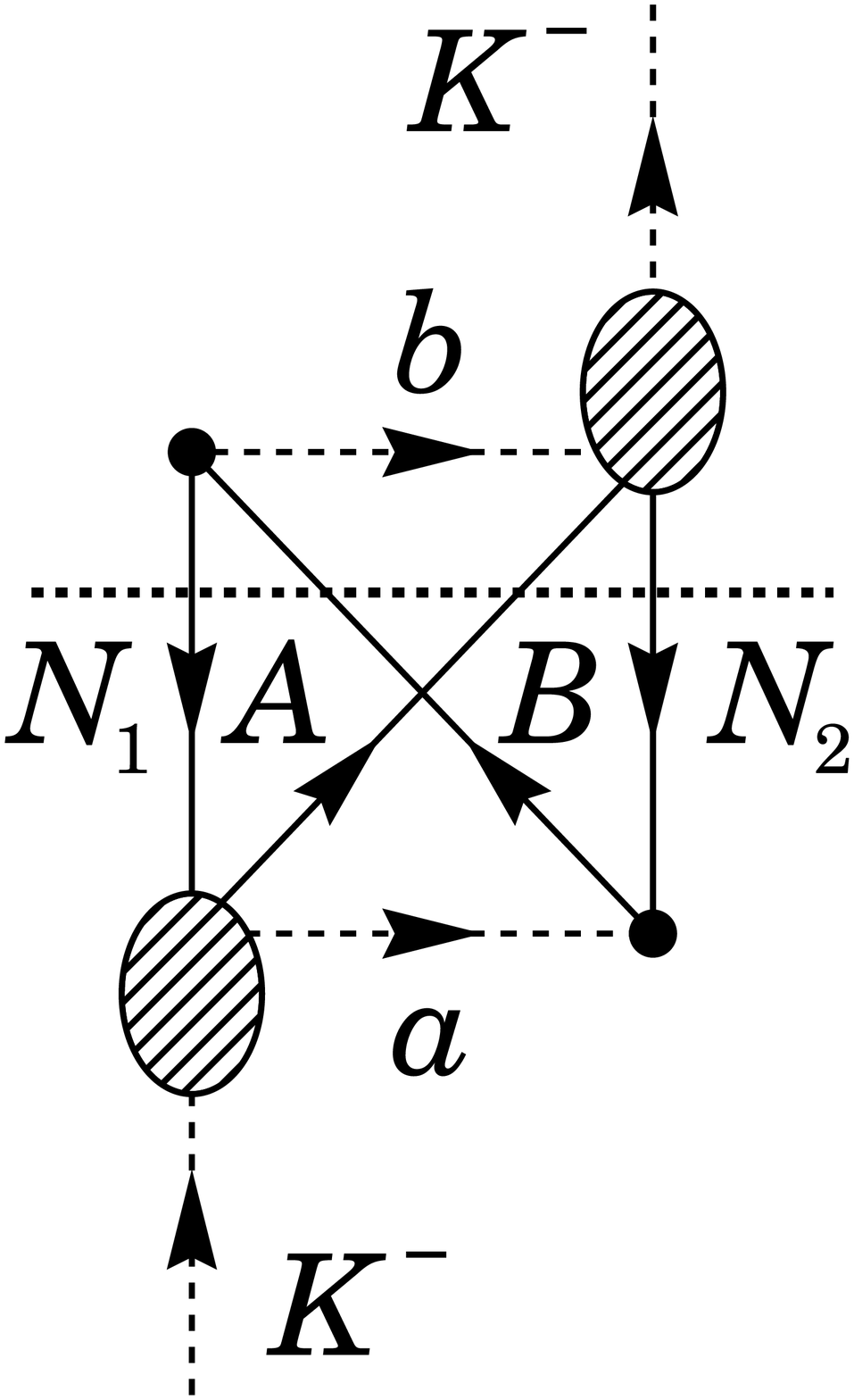}
    \\ (a) & (b) & (c) & (d) & (e)
  \end{tabular*}
  \caption{Feynman diagrams for the mesonic (a) and non-mesonic
    (b--e) $K^{-}$ absorption processes in nuclear
    matter~\cite{Sekihara:2012wj}.  In the mesonic diagram, $N_{1}$,
    $\pi$, and $Y$ denote nucleons, pions, and hyperons,
    respectively. In the non-mesonic diagrams, $N_{1}$ and $N_{2}$
    denote nucleons, $A$ and $B$ baryons, and $a$ and $b$ mesons.
    The shaded ellipses represent the $K^{-}N \to MB$ amplitudes. }
  \label{fig1}
\end{figure}

We evaluate the absorption potential for $K^{-}$ in nuclear matter as
the imaginary part of the $K^{-}$ self-energy in nuclear matter as a
function of nuclear density $\rho _{N}$.  The mesonic absorption
potential is obtained by considering the self-energy with one-body
process as shown in Fig.~\ref{fig1}(a).  For the non-mesonic
absorption, we calculate two-body absorption with one-meson exchange
model where the Nambu-Goldstone bosons are exchanged between the
baryons, as shown in Figs.~\ref{fig1}(b--e).  In the calculation, the
$K^{-}N \to MB$ scattering amplitudes are determined by using the
so-called chiral unitary approach~\cite{Kaiser:1995eg}, which is based
on chiral dynamics and a unitarized framework.  In the chiral unitary
approach the low-energy $\bar{K}N$ scattering is well reproduced and
$\Lambda (1405)$ is dynamically generated.  We describe nuclear matter
using the Thomas-Fermi approximation, in which a bound nucleon with
momentum $p$ has energy, $E_{N} = M_{N} + p^{2} / (2 M_{N}) - k_{\rm
  F}^{2}/(2 M_{N})$, with the nucleon mass $M_{N}$ and the Fermi
momentum $k_{\rm F}= (3 \pi ^{2} \rho _{N}/2)^{1/3}$.  The details of
the formulation for the absorption potential are given in
Ref.~\cite{Sekihara:2012wj}.

One important feature of $K^{-}$ absorption in nuclei is that the
center-of-mass energy of the $K^{-}N$ pair in the initial state can go
below the threshold value, $m_{\bar{K}} + M_{N}$, where $m_{\bar{K}}$
is the in-vacuum $K^{-}$ mass, due to the off-shellness of the bound
nucleon.  For a kaon with a finite momentum and a binding energy, the
two-body energy shifts farther downward compared to the kaon at rest
owing to the off-shellness of the kaon.  Furthermore, the $K^{-}N$
system in the initial state can have lower energies in higher
densities.  These facts mean that strength of the contributions from
$\Lambda (1405)$, existing below the $K^{-} p$ threshold, to the
absorption reactions depends on the nuclear density as well as the
kaon energy and momentum.  In our model setup, the $K^{-}p$ energy
with $K^{-}$ at rest becomes around $1420 \text{ MeV}$ at the nuclear
density $\rho _{N} \approx 0.05$--$0.1 \text{ fm}^{-3}$.  This energy
corresponds to the peak position of the $\Lambda (1405)$ spectra in
$K^{-}p \to (\pi \Sigma) ^{0}$.  Even at the saturation density $\rho
_{0}=0.17 \text{ fm}^{-3}$ the $K^{-}p$ energy is around $1400 \text{
  MeV}$, which is in the $\Lambda (1405)$ peak.

\begin{figure}[t]
  \centering
    \includegraphics[width=7.5cm]{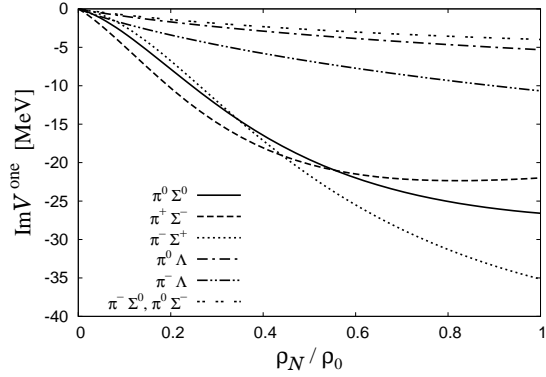}
    \caption{Mesonic absorption potential ($\text{Im} V^{\text{one}}$)
      for $K^{-}$ at rest in nuclear matter as a function of nuclear
      density $\rho _{N}$~\cite{Sekihara:2012wj}.  The potentials for
      $K^{-} n \to \pi ^{-} \Sigma ^{0}$ and $\pi ^{0} \Sigma ^{-}$
      have the same values owing to the isospin symmetry.  }
  \label{fig:abs-one}
\end{figure}

We now show in Fig.~\ref{fig:abs-one} the mesonic absorption potential
for $K^{-}$ at rest in nuclear matter as a function of nuclear density
$\rho _{N}$.  As one can see from the figure, the absorption to the $(\pi
\Sigma )^{0}$ states is dominant to the other channels.  Bearing in
mind that the $\Lambda (1405)$ appears selectively in the $K^{-}p \to
(\pi \Sigma )^{0}$ transitions, this result shows that the $\Lambda
(1405)$ contribution is important for the mesonic absorption of $K^{-}$
in these densities and $\Lambda (1405)$ can be a doorway to the
absorption of $K^{-}$ at rest.  We also note that the potential for
the $(\pi \Sigma )^{0}$ channels does not show $\rho _{N}^{1}$-like
dependence owing to the energy dependence of the $\bar{K}N$ amplitude
coming from the $\Lambda (1405)$ resonance.  Another interesting
feature for the mesonic absorption is that the behaviors of the
absorption to $\pi ^{\pm} \Sigma ^{\mp}$ are different from each
other; while the ratio $R_{+-} \equiv [\pi ^{-} \Sigma ^{+}] / [\pi
^{+} \Sigma ^{-}]$ is less than unity at $\rho _{N} < 0.08 \text{
  fm}^{-3}$, it gets larger as the density increases and becomes about
$1.6$ at the saturation density.  This tendency comes from the facts
that the $\Lambda (1405)$ peak in the $K^{-}p \to \pi ^{+} \Sigma
^{-}$ ($\pi ^{-} \Sigma ^{+}$) shifts upward (downward) due to the
interference between $\Lambda (1405)$ and the $I=1$ non-resonant
background and the lower (higher) density probes the higher (lower)
energy of the initial $K^{-}p$ system.  We note that the behavior of the
ratio $R_{+-}$ is consistent with the experimental results.  Namely,
while the ratio $R_{+-}$ is $0.42$ for stopped $K^{-}$ on
proton~\cite{Tovee:1971ga}, which constrains the ratio at zero density
in our model, it becomes large as off-shellness of bound nucleons
increases, as $0.85$ on deuteron and $1.8 \pm 0.5$ on ${}^{4}
\text{He}$~\cite{Katz:1970ng}, and $1.2$--$1.5$ on $p$-shell
nuclei~\cite{Agnello:2011iq}.  This indicates that the experimental
results on $R_{+-}$ of the stopped $K^{-}$ absorption in various nuclei
could be explained by the nature of the $\Lambda (1405)$ resonance
together with the $I=1$ non-resonant background.  We also note that
the condition of $R_{+-}=1$ takes place at relatively lower density,
$\rho _{N}\approx 0.08 \text{ fm}^{-3}$, which means that the peak
position of the $\Lambda (1405)$ spectrum in $K^{-}p \to (\pi \Sigma
)^{0}$ should be at an energy close to the $\bar{K}N$ threshold rather
than at $1405 \text{ MeV}$.

\begin{figure}[!b]
  \centering
  \begin{tabular*}{\textwidth}{@{\extracolsep{\fill}}cc}
    \includegraphics[width=7.5cm]{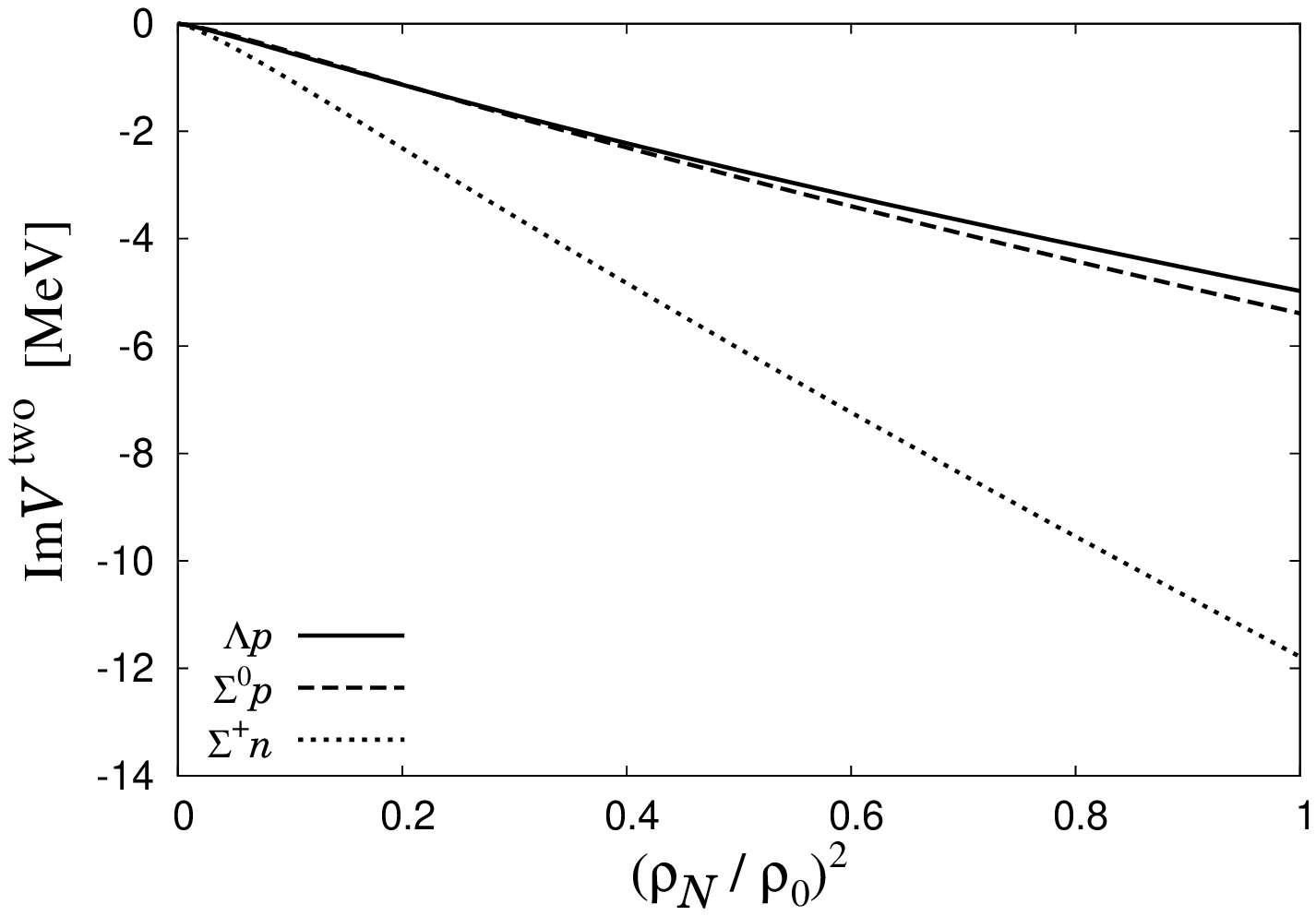} & 
    \includegraphics[width=7.5cm]{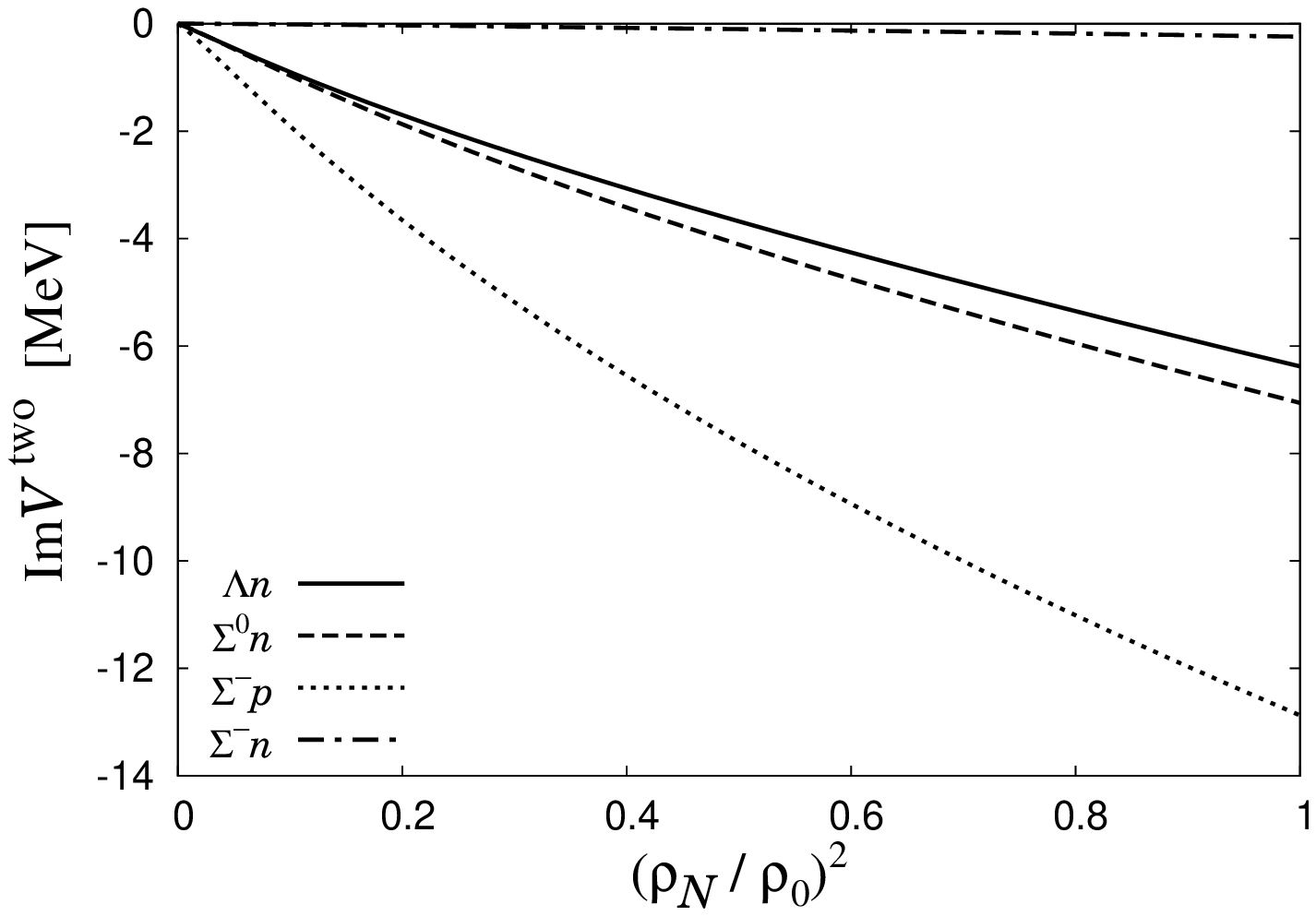} 
  \end{tabular*}
  \caption{Non-mesonic absorption potential ($\text{Im}
    V^{\text{two}}$) for $K^{-}$ at rest in nuclear matter as a
    function of nuclear density squared $\rho
    _{N}^{2}$~\cite{Sekihara:2012wj}.  Left (right) panel shows the
    contributions from the $K^{-}pp \to \Lambda p$, $\Sigma ^{0} p$,
    and $\Sigma ^{+} n$ processes (the $K^{-}pn \to \Lambda n$,
    $\Sigma ^{0} n$, $\Sigma ^{-} p$ and $K^{-}nn \to \Sigma ^{-} n$
    processes). }
  \label{fig:abs-two} 
\end{figure}%

Next we show in Fig.~\ref{fig:abs-two} the non-mesonic absorption
potential for $K^{-}$ at rest in nuclear matter as a function of
nuclear density squared.  From the figure, we find large contributions from
the $K^{-}pp$ and $K^{-}pn$ initial state to the non-mesonic
absorption while tiny absorption potential from the $K^{-}nn$ initial
state.  Considering that $K^{-}$ absorption with a proton induces the
$\Lambda (1405)$ resonance, we can see that the $\Lambda (1405)$
doorway process is realized also in non-mesonic absorption in the
$K^{-}pp$ and $K^{-}pn$ contribution.  We also note that the potential
shows non-$\rho _{N}^{2}$ dependence due to the $\Lambda (1405)$
existence.  
The deviation from $\rho _{N}^{2}$ dependence is most significant in
the $K^{-} p n \to \Sigma ^{-} p$ process at lower densities, because
in this process $K^{-}p \to \pi ^{+} \Sigma ^{-}$ transition
dominantly takes place, in which the transition strength is maximum at
$1424 \text{ MeV}$ from the $\Lambda (1405)$ resonance and the peak
position is higher than the other channels as mentioned in the mesonic
absorption.
Furthermore, we have the absorption ratios $[\Lambda p] /
[\Sigma ^{0} p] \approx [\Lambda n] / [\Sigma ^{0} n] \approx 1$ and
$[\Sigma ^{+} n] / [\Sigma ^{0} p] \approx [\Sigma ^{-} p] / [\Sigma
^{0} n] \approx 2$ with marginal dependence on the nuclear density.
Since in Ref.~\cite{Sekihara:2009yk} the non-mesonic decay of $\Lambda
(1405)$ in nuclear matter is found to be $[\Lambda N] / [\Sigma ^{0}
N] \approx 1.2$ in the chiral unitary approach and $[\Sigma ^{\pm} N]
/ [\Sigma ^{0} N]$ is exactly two if the $\bar{K} N$ is purely $I=0$,
the present result of the non-mesonic absorption ratios can be
interpreted as the consequence of the $\Lambda (1405)$ dominance.

We now summarize the results of the mesonic and non-mesonic fractions
of the $K^{-}$ absorption at rest.  It is found that the mesonic
(non-mesonic) fraction almost linearly goes down (up) from unity
(zero) as the nuclear density increases, because the non-mesonic
reaction can more largely contribute to the absorption at higher
densities.  At the saturation density the mesonic and non-mesonic
fractions are about $70 \%$ and $30 \%$, respectively.  These
fractions are close to the empirical value for kaonic atoms with
nuclei heavier than ${}^{4} \text{He}$ (about $80 \%$ and $20 \%$,
respectively~\cite{Friedman:2007zza}).  
However, it is suggested in Refs.~\cite{Yamagata-Sekihara:HYP,
  Gal:2013vx} that the effective density where the absorption mainly
takes place does not correspond to the nuclear saturation density but
to a fraction of it and the effective density strongly depends on the
strong interaction between $K^{-}$ and nucleus.  Therefore, bearing in
mind that for bound kaons the kaon momenta and energies are determined
self-consistently in the equation of motion with the
energy-momentum-dependent potential, a realistic treatment of the
bound kaon with finite nuclei is necessary for quantitative
discussions.



Finally we calculate the branching ratios of the $K^{-}$ absorption at
rest in deuteron and ${}^{4} \text{He}$ by using phenomenological
wave functions for bound nucleons~\cite{Sekihara:next}.  These light
nuclei serve as environments of various nuclear densities inside
nuclei due to the large varieties of the binding energies per one
nucleon.  The $R_{+-}$ ratio is found to be 0.84 
and 1.32 
for deuteron and ${}^{4} \text{He}$ targets, respectively.  The
non-mesonic fraction amounts to $3.4 \%$ in the $K^{-}$-deuteron case
and to $18.6 \%$ in the $K^{-}$-${}^{4} \text{He}$ case.  Both results
are close to the experimental results.  For quantitative comparison
with experimental results, however, we have to take into account final
state interactions such as $\Sigma$-$\Lambda$ conversions, and also
appropriate wave function of bound $K^{-}$ will be necessary because
$K^{-}$ even in atomic states may have a large momentum inside nucleus
due to the orthogonality to the kaonic nuclear
states~\cite{Cieply:2011yz}.

\section{Summary}
\label{sec3}

In this study we have investigated the mesonic and non-mesonic
branching ratios of $K^{-}$ in nuclear matter from the $K^{-}$
self-energy as functions of nuclear density.  By using the chiral
unitary approach for the $s$-wave $\bar{K} N$ interaction as an input,
we have found that both the mesonic and non-mesonic absorptions of
$K^{-}$ at rest are dominated by the $\Lambda (1405)$ contribution.
The absorption ratio $R_{+-}\equiv [\pi ^{-} \Sigma ^{+}] / [\pi ^{+}
\Sigma ^{-}]$ becomes larger as the nuclear density increases and its
behavior is consistent with the experimental results.  The ratio
$R_{+-}$ becomes unity at relatively lower density, which means that
the peak position of the $\Lambda (1405)$ spectrum in $K^{-}p \to (\pi
\Sigma )^{0}$ should be at an energy close to the $\bar{K}N$ threshold
rather than at $1405 \text{ MeV}$.  The non-mesonic absorption ratios
$[\Lambda p]/ [\Sigma ^{0} p]$ and $[\Lambda n]/ [\Sigma ^{0} n]$ are
about unity while $[\Sigma ^{+}n]/[\Sigma ^{0}p]$ and $[\Sigma
^{-}p]/[\Sigma ^{0}n]$ are about two due to the $\Lambda (1405)$
dominance in absorption.  The non-mesonic fraction is about $30 \%$ at
the saturation density.  We have also calculated the branching ratios
of the $K^{-}$ absorption at rest in deuteron and ${}^{4}
\text{He}$, and we have found that the evaluated $R_{+-}$ ratio and
the non-mesonic fraction are close to the experimental results.







\end{document}